\documentstyle[12pt]{article}
\advance \topmargin by -\headheight
\advance \topmargin by -\headsep

\evensidemargin \oddsidemargin
\begin{document}

\topmargin -.6in

\def\rf#1{(\ref{eq:#1})}
\def\lab#1{\label{eq:#1}}
\def\nonu{\nonumber}
\def\br{\begin{eqnarray}}
\def\er{\end{eqnarray}}
\def\be{\begin{equation}}
\def\ee{\end{equation}}
\def\eq{\!\!\!\! &=& \!\!\!\! }
\def\foot#1{\footnotemark\footnotetext{#1}}
\def\lb{\lbrack}
\def\rb{\rbrack}
\def\llangle{\left\langle}
\def\rrangle{\right\rangle}
\def\blangle{\Bigl\langle}
\def\brangle{\Bigr\rangle}
\def\llb{\left\lbrack}
\def\rrb{\right\rbrack}
\def\Blb{\Bigl\lbrack}
\def\Brb{\Bigr\rbrack}
\def\lcurl{\left\{}
\def\rcurl{\right\}}
\def\({\left(}
\def\){\right)}
\def\v{\vert}                     %% vertical bars
\def\bv{\bigm\vert}               %%
\def\Bgv{\;\Bigg\vert}            %%
\def\bgv{\bigg\vert}              %%
\def\lskip{\vskip\baselineskip\vskip-\parskip\noindent}
\def\mskp{\par\vskip 0.3cm \par\noindent}
\def\sskp{\par\vskip 0.15cm \par\noindent}
\relax

\def\tr{\mathop{\rm tr}}                  % tr - small trace
\def\Tr{\mathop{\rm Tr}}                  % Tr - big trace
\newcommand\partder[2]{{{\partial {#1}}\over{\partial {#2}}}}
                                                  % partial derivative
\newcommand\funcder[2]{{{\delta {#1}}\over{\delta {#2}}}}
                                                % functional derivative
\newcommand\Bil[2]{\Bigl\langle {#1} \Bigg\vert {#2} \Bigr\rangle}  %% <.|.>
\newcommand\bil[2]{\left\langle {#1} \bigg\vert {#2} \right\rangle} %% <.|.>
\newcommand\me[2]{\langle {#1}\vert {#2} \rangle} %% <.|.>

\newcommand\sbr[2]{\left\lbrack\,{#1}\, ,\,{#2}\,\right\rbrack} % commutator
\newcommand\Sbr[2]{\Bigl\lbrack\,{#1}\, ,\,{#2}\,\Bigr\rbrack} % commutator
%%(Large)
\newcommand\pbr[2]{\{\,{#1}\, ,\,{#2}\,\}}       % Poisson brackets
\newcommand\Pbr[2]{\Bigl\{ \,{#1}\, ,\,{#2}\,\Bigr\}}  % Poisson brackets
%%(large)
\newcommand\pbbr[2]{\lcurl\,{#1}\, ,\,{#2}\,\rcurl}  % Poisson brackets
%%(left-right)
\newcommand\sumi[1]{\sum_{#1}^{\infty}}   %% summation till infinity

\def\a{\alpha}
\def\b{\beta}
\def\c{\chi}
\def\d{\delta}
\def\D{\Delta}
\def\eps{\epsilon}
\def\vareps{\varepsilon}
\def\g{\gamma}
\def\G{\Gamma}
\def\grad{\nabla}
\def\h{{1\over 2}}
\def\k{\kappa}
\def\l{\lambda}
\def\L{\Lambda}
\def\m{\mu}
\def\n{\nu}
\def\om{\omega}
\def\O{\Omega}
\def\p{\phi}
\def\P{\Phi}
\def\pa{\partial}
\def\pr{\prime}
\def\ra{\rightarrow}
\def\lra{\longrightarrow}
\def\s{\sigma}
\def\S{\Sigma}
\def\t{\tau}
\def\th{\theta}
\def\Th{\Theta}
\def\z{\zeta}
\def\ti{\tilde}
\def\wti{\widetilde}
\def\one{\hbox{{1}\kern-.25em\hbox{l}}}

\def\cA{{\cal A}}
\def\cB{{\cal B}}
\def\cC{{\cal C}}
\def\cD{{\cal D}}
\def\cE{{\cal E}}
\def\cH{{\cal H}}
\def\cJ{{\cal J}}
\def\cL{{\cal L}}
\def\cM{{\cal M}}
\def\cN{{\cal N}}
\def\cP{{\cal P}}
\def\cQ{{\cal Q}}
\def\cR{{\cal R}}
\def\cS{{\cal S}}
\def\cT{{\cal T}}
\def\cU{{\cal U}}
\def\cV{{\cal V}}
\def\cW{{\cal W}}
\def\cY{{\cal Y}}

\def\phanta{\phantom{aaaaaaaaaaaaaaa}}
\def\phantb{\phantom{aaaaaaaaaaaaaaaaaaaaaaaaa}}
\def\phantc{\phantom{aaaaaaaaaaaaaaaaaaaaaaaaaaaaaaaaaaa}}

\def\symp#1{{\cal S}{\cal D}if\!\! f \, ({#1})}
\def\esymp#1{{\wti {\cal S}{\cal D}if\!\! f} \, ({#1})}
\def\Symp#1{{\rm SDiff}\, ({#1})}
\def\eSymp#1{{\wti {\rm SDiff}}\, ({#1})}
\def\vol#1{{{\cal D}if\!\! f}_0 ({#1})}
\def\Vol#1{{\rm Diff}_0 ({#1})}

\def\Winf{{\bf W_\infty}}               % Linear W-infinity
\def\Win1{{\bf W_{1+\infty}}}           % Linear W-1+infinity
\def\nWinf{{\bf {\hat W}_\infty}}       % Nonlinear W-infinity
\def\PsDA{\Psi{\cal DO}}
                   % algebra of all pseudo-differential operators

%%
\newcommand{\nit}{\noindent}
\newcommand{\ct}[1]{\cite{#1}}
\newcommand{\bi}[1]{\bibitem{#1}}
\newcommand\PRL[3]{{\sl Phys. Rev. Lett.} {\bf#1} (#2) #3}
\newcommand\NPB[3]{{\sl Nucl. Phys.} {\bf B#1} (#2) #3}
\newcommand\NPBFS[4]{{\sl Nucl. Phys.} {\bf B#2} [FS#1] (#3) #4}
\newcommand\CMP[3]{{\sl Commun. Math. Phys.} {\bf #1} (#2) #3}
\newcommand\PRD[3]{{\sl Phys. Rev.} {\bf D#1} (#2) #3}
\newcommand\PLA[3]{{\sl Phys. Lett.} {\bf #1A} (#2) #3}
\newcommand\PLB[3]{{\sl Phys. Lett.} {\bf #1B} (#2) #3}
\newcommand\JMP[3]{{\sl J. Math. Phys.} {\bf #1} (#2) #3}
\newcommand\PTP[3]{{\sl Prog. Theor. Phys.} {\bf #1} (#2) #3}
\newcommand\SPTP[3]{{\sl Suppl. Prog. Theor. Phys.} {\bf #1} (#2) #3}
\newcommand\AoP[3]{{\sl Ann. of Phys.} {\bf #1} (#2) #3}
\newcommand\RMP[3]{{\sl Rev. Mod. Phys.} {\bf #1} (#2) #3}
\newcommand\PR[3]{{\sl Phys. Reports} {\bf #1} (#2) #3}
\newcommand\FAP[3]{{\sl Funkt. Anal. Prilozheniya} {\bf #1} (#2) #3}
\newcommand\FAaIA[3]{{\sl Functional Analysis and Its Application} {\bf #1}
(#2) #3}
\def\InvM#1#2#3{{\sl Invent. Math.} {\bf #1} (#2) #3}
\newcommand\LMP[3]{{\sl Letters in Math. Phys.} {\bf #1} (#2) #3}
\newcommand\IJMPA[3]{{\sl Int. J. Mod. Phys.} {\bf A#1} (#2) #3}
\newcommand\TMP[3]{{\sl Theor. Mat. Phys.} {\bf #1} (#2) #3}
\newcommand\JPA[3]{{\sl J. Physics} {\bf A#1} (#2) #3}
\newcommand\JSM[3]{{\sl J. Soviet Math.} {\bf #1} (#2) #3}
\newcommand\MPLA[3]{{\sl Mod. Phys. Lett.} {\bf A#1} (#2) #3}
\newcommand\JETP[3]{{\sl Sov. Phys. JETP} {\bf #1} (#2) #3}
\newcommand\JETPL[3]{{\sl  Sov. Phys. JETP Lett.} {\bf #1} (#2) #3}
\newcommand\PHSA[3]{{\sl Physica} {\bf A#1} (#2) #3}
\newcommand\PHSD[3]{{\sl Physica} {\bf D#1} (#2) #3}

\newcommand\hepth[1]{{\sl hep-th/#1}}

\begin{titlepage}
\vspace*{-1cm}
\noindent
May, 1995 \hfill{BGU-95 / 10 / May - PH}\\
\phantom{bla}
\hfill{hep-th/9505128}
\\
\begin{center}
{\large {\bf Volume-Preserving Diffeomorphisms' versus Local Gauge Symmetry}}
\end{center}
\vskip .3in
\begin{center}
E.I. Guendelman, E. Nissimov$^{\, 1,2}$  and
S. Pacheva\foot{On leave from the Institute of Nuclear Research and Nuclear
Energy, Boul. Tsarigradsko Chausee 72, BG-1784 ~Sofia, Bulgaria;
{\em e-mail}: emil@bgearn.bitnet, svetlana@bgearn.bitnet}
\foot{Supported in part by Bulgarian NSF grant {\em Ph-401}.} \\
Department of Physics, Ben-Gurion University of the Negev \\
Box 653, IL-84105 ~Beer Sheva, Israel \\
{\small {\em e-mail}: guendel@bguvm.bgu.ac.il, emil@bgumail.bgu.ac.il,
svetlana@bgumail.bgu.ac.il}
\end{center}
\vskip .3in

\begin{abstract} \noindent
We present a new form of Quantum Electrodynamics where the photons are
composites made out of zero-dimensional scalar ``primitives''. The r\^{o}le of
the local gauge symmetry is taken over by an {\em infinite-dimensional global
Noether symmetry} -- the group of volume-preserving (symplectic)
diffeomorphisms
of the target space of the scalar primitives. Similar construction is carried
out for higher antisymmetric tensor gauge theories. Solutions of Maxwell's
equations are automatically solutions of the new system. However, the latter
possesses additional non-Maxwell solutions which display some
interesting new effects: (a) a magneto-hydrodynamical analogy,
(b) absence of electromagnetic self-energy
for electron plane wave solutions, and (c) gauge invariant photon mass
generation, where the generated mass is arbitrary.
\end{abstract}

\end{titlepage}

\noindent
{\large {\bf 1. Introduction}}
\mskp

Systems with infinite number of conservation laws have been extensively
studied,
because their high symmetry allows to extract non-perturbative information and
in some instances, {\sl e.g.} for the completely integrable two-dimensional
field-theoretic models, even to solve them exactly \ct{integr}. It is however
hard to find a realistic field theory in $D=4$ space-time dimensions,
that allows infinite number of nontrivial conserved charges.

In this work, we will study a theory which allows an infinite number
of conservation laws and which is at the same time a theory resembling very
much the most respectable and most tested field theory that we know up to date
-- Quantum Electrodynamics (QED).
This is achieved through the substitution of the ordinary local gauge symmetry
of QED (which does not lead to conservation laws, but rather to
constraints on the physical degrees of freedom) by a {\em global
infinite-dimensional} Noether symmetry group -- the group of volume-preserving
(symplectic) diffeomorphisms (see, {\sl e.g.} \ct{Arnold}).
An infinite-dimensional abelian group, corresponding to the Cartan
subalgebra of the group of volume-preserving diffeomorphisms, has already
been identified as a symmetry of a variant of the model studied here by one of
us \ct{mini-QED}.

The above mentioned Noether symmetry group acts as a group of transformations
on a set of primitive dimensionless scalar fields taking values in some smooth
manifold (``target'' space)  and preserves the volume form there.
The gauge potential and the gauge field strength (the photons) in such a theory
are not fundamental fields, but rather they are composites of these primitive
scalar constituents. Under a volume-preserving diffeomorphism transformation,
the composite gauge field is transformed by the addition of a total gradient
while the composite field strength is, of course, a
volume-preserving-diffeomorphism-invariant quantity.
It is possible to couple the composite gauge field to matter fields in the
well known minimal way so that the gradient, that is added to the gauge field
under a volume-preserving transformation on the scalar constituents,
is canceled by an appropriate phase transformation of the matter fields.
If these matter fields are taken to be Dirac fermion fields,
we obtain a ``mini-QED'' model that resembles very much the usual QED
describing the interaction of electromagnetic field with Dirac particles.
Furthermore, if the number
of the primitive scalars is greater or equal to $D$ (dimension of space-time),
any solution of the usual Maxwell's equations  can be expressed (in a
non-unique way, in general) through the primitive scalars and it is
automatically a solution of mini-QED equations of motion.

However the mini-QED model does not coincide completely with QED, since it
possesses solutions that do not respect Maxwell's equations. These additional
solutions describe new effects that do not appear in the usual formulation of
electrodynamics. Among these new features
we can notice at least the following: (a) a magneto-hydrodynamical analogy,
(b) absence of electromagnetic self-energy
for electron plane wave solutions, and (c) generation of gauge invariant
photon mass.
In addition, as it was already observed in a particular variant of the mini-QED
model \ct{mini-QED}, the idea of substituting local gauge invariance with a
global infinite-dimensional symmetry group can be applied to the
construction of models with chiral invariance which are totally anomaly free
(provided the number primitive scalars is smaller than the dimension of
space-time).

The above construction is carried out here also for higher antisymmetric tensor
(``p-form'') gauge theories (see, {\sl e.g.} \ct{p-form}),
for instance -- the Kalb-Ramond ``2-form'' gauge model \ct{K-R},
where the ordinary local ``p-form'' tensor gauge invariance is substituted with
the {\em global infinite-dimensional} Noether symmetry of
volume-preserving diffeomorphisms on the $p+1$-dimensional target space of
primitive scalar constituents. Here again, in general, all solutions of the
ordinary tensor gauge theories are automatically solutions (as composites
made out of the primitive scalar constituents) to the ``mini-tensor-gauge''
theories, however, the latter possess additional solutions absent from the
usual $p$-form gauge models.
\lskip
\noindent
{\large {\bf 2. Groups of Volume-Preserving Diffeomorphisms}}
\mskp
First, let us recall the basic notions connected with groups of symplectic
diffeomorphisms and more generally -- the groups of volume-preserving
diffeomorphisms on smooth manifolds (see, {\sl e.g.} \ct{Arnold}).
Let $\cT^{2n}$ be a ($2n$-dimensional) symplectic manifold with a symplectic
structure which can always (at least locally) be
represented in terms of a canonical constant anti-symmetric $2n \times 2n$
matrix:
\be
\Vert \om_{ab} \Vert =
\left(\begin{array}{ccccc}
0 & 1 & 0 & 0 & \ldots \\
-1 & 0 & 0 & 0 & \ldots \\
0 & 0 & 0 & 1 & \ldots \\
0 & 0 & -1 & 0 & \ldots \\
\vdots & \vdots & \vdots & \vdots & \ddots
\end{array} \right)
\lab{omega}
\ee
and let $\lcurl \P^a \rcurl_{a=1}^{2n}$ denote the corresponding (local)
coordinates on $\cT^{2n}$. Then the infinite-dimensional group
$\Symp{\cT^{2n}}$ of
symplectic diffeomorphisms $\P^a \to G^a \( \P\)$ on the manifold $\cT^{2n}$
and the associated infinite-dimensional Lie algebra $\symp{\cT^{2n}}$ of
infinitesimal symplectic diffeomorphisms
$G^a \(\P\) \approx \P^a + \om^{ab} \partder{\G}{\P^b}$ are defined as follows:
\br
\Symp{\cT^{2n}} \equiv \lcurl  \P^a \to G^a \( \P\) \; ;\;
\om_{cd} \partder{G^c}{\P^a} \partder{G^d}{\P^b} = \om_{ab} \rcurl
\lab{symp-diff-group}  \\
\symp{\cT^{2n}} \equiv \lcurl \G \(\P\) \; ;\; \sbr{\G_1}{\G_2} \equiv
\pbbr{\G_1}{\G_2} = \om^{ab} \partder{\G_1}{\P^a} \partder{\G_2}{\P^b} \rcurl
\lab{symp-diff-alg}
\er
As seen from eq.\rf{symp-diff-alg}, the Lie-commutator is nothing but the
canonical Poisson bracket on $\cT^{2n}$ with $\om^{ab}$ indicating the inverse
matrix w.r.t. $\om_{ab}$ \rf{omega}.

In the simplest case $n=1$, $\om_{ab} = \vareps_{ab} \; ,\vareps_{12}=1$, and
$\symp{\cT^{2}}$ is precisely the algebra of area-preserving diffeomorphisms on
a two-dimensional manifold $\cT^2$, also known as $w_{\infty}$-algebra
when $\cT^2$ is a cylinder (for a review, see \ct{W-inf} and references
therein), which contains as a subalgebra the centerless
conformal Virasoro algebra. For example, in the case of torus
$\cT^2 = S^1 \times S^1$ the Lie-algebra elements of $\symp{S^1 \times S^1}$
are
given by:
\be
\G \(\P\) = \sum_{\vec{n}} \g_{\vec{n}} \G_{\vec{n}}\(\P\) \qquad , \quad
\G_{\vec{n}}\(\P\) \equiv \exp \lcurl \vec{n}.\vec{\Phi} \rcurl \;\; , \;
\vec{n} = \( n_1 ,n_2\)
\lab{torus}
\ee
where $\( n_1 ,n_2 \)$ are arbitrary integers.
Furthermore, $\symp{S^1 \times S^1}$ allows for a non-trivial central
extension \ct{Floratos} which in the basis $\lcurl \G_{\vec{n}} \rcurl$ reads:
\be
\Sbr{\G_{\vec{n}}}{\G_{\vec{m}}} = -\( n_1 m_2 - n_2 m_1\) \G_{\vec{n}+\vec{m}}
- \vec{c}.\vec{n} \d_{\vec{n} + \vec{m},0}
\lab{ext-torus-alg}
\ee
where $\vec{c}=\( c_1 ,c_2\)$ denotes the pair of ``central charges''.

It is well known that the group $\Symp{\cT^{2n}}$ \rf{symp-diff-group} is a
subgroup of the group $\Vol{\cT^{2n}}$ of {\em volume-preserving}
diffeomorphisms
on the manifold $\cT^{2n}$ . Moreover, the group $\Vol{\cT^{s}}$ exists for
manifolds of arbitrary (not necessarily even) dimension $s$ :
\be
\Vol{\cT^{s}} \equiv \lcurl  \P^a \to G^a \(\P\) \; ;\;
\vareps_{b_1 ,\ldots ,b_s} \partder{G^{b_1}}{\P^{a_1}} \cdots
\partder{G^{b_s}}{\P^{a_s}} = \vareps_{a_1 ,\ldots ,a_s}  \rcurl
\lab{vol-diff-group}
\ee
Indeed, it is straightforward to verify that
$G^a \(\P\)$ defined in \rf{symp-diff-group} preserves the volume-form
$\frac{1}{s!} \vareps_{a_1 \ldots a_s} d\P^{a_1} \wedge \ldots \wedge
d\P^{a_s}$ on ${\cT^{s}}$ where the simple relation was used (for $s=2n$) :
\be
\om_{\lb a_1 a_2} \cdots \om_{a_{s-1} a_{s}\rb} =
\vareps_{a_1 \ldots a_{s}}
\lab{omega-eps}
\ee
and the square brackets indicate total antisymmetrization of indices.
Accordingly, the Lie algebra $\vol{\cT^{s}}$ of infinitesimal volume-preserving
diffeomorphisms is given by:
\br
\vol{\cT^{s}} \equiv \lcurl \G^a \(\P\) \;\; ;\;\;
G^a \(\P\) \approx \P^a + \G^a \(\P\) \; ,\; \partder{\G^b}{\P^b} =0 \rcurl
\lab{vol-diff-alg} \\
i.e. \quad  \G^a \(\P\) = \frac{1}{(s-2)!} \vareps^{a b c_1 \ldots c_{s-2}}
\partder{}{\P^b}\G_{c_1 \ldots c_{s-2}}
\lab{rot}
\er
\lskip
{\large {\bf 3. Field-Theory Models}}
\mskp
Now, let us consider a set of $2n$ (zero-dimensional) scalar fields
$\lcurl \P^a (x) \rcurl_{a=1}^{2n}$ on ordinary Minkowski space-time
taking values in the
symplectic manifold $\cT^{2n}$. Then, the canonical symplectic closed two-form
$\O$ on $\cT^{2n}$ naturally leads to the construction of an antisymmetric
tensor field $F_{\m\n} \( \P (x)\)$ on space-time satisfying the Bianchi
identity and, therefore, allowing for a potential $A_\m \(\P (x)\)$ :
\br
\O = \h \om_{ab} d\P^a \wedge d\P^b = \h F_{\m\n} \( \P\) dx^\m \wedge dx^\n
\lab{diff-form}  \\
F_{\m\n} \( \P\) = \om_{ab} \pa_\m \P^a \pa_\n \P^b   \quad ,\quad
\pa_\k F_{\l\m}\(\P\) + \pa_\m F_{\k\l}\(\P\)  + \pa_\l F_{\m\k}\(\P\) = 0
\lab{F-primitiv} \\
F_{\m\n} \( \P\) = \pa_\m A_\n \( \P\) - \pa_\n A_\m \( \P\) \qquad ,\qquad
A_\m \( \P\) = \h \om_{ab} \P^a \pa_\m \P^b
\lab{A-primitiv}
\er
Clearly, in eq.\rf{A-primitiv} $A_\m \( \P\)$ is determined up to a
$\P$-dependent total derivative.

{}From the basic definitions \rf{symp-diff-group} and \rf{symp-diff-alg} one
immediately finds that the field $F_{\m\n} \( \P\)$ is invariant under
arbitrary
field transformations (reparametrizations) $\P^a (x) \to G^a \(\P (x)\)$
belonging to the infinite-dimensional group $\Symp{\cT^{2n}}$, whereas its
potential transforms with a $\P$-dependent total derivative:
\be
F_{\m\n} \( G\(\P\)\)  = F_{\m\n} \( \P\)  \quad ,\quad
A_\m \( \P\) \to  A_\m \( \P\) + \pa_\m \( \G\(\P\) -
\h \P^a \partder{\G}{\P^a}\)
\lab{transform}
\ee

Eqs.\rf{F-primitiv}--\rf{transform} naturally suggest the interpretation of
$F_{\m\n} \( \P\)$ as an electromagnetic field strength and $A_\m \( \P\)$ as
the corresponding vector potential which are now {\em composite} fields made
out of more elementary ``primitive'' zero-dimensional scalar fields $\P^a (x)$.
Accordingly, the r\^{o}le of the ordinary local gauge invariance is now taken
over by the infinite-dimensional $\Symp{\cT^{2n}}$ {\em global Noether}
symmetry
\rf{symp-diff-group}.

Thus, we can consider the following model of scalar ``primitives'' $\P^a (x)$
coupled to ordinary Dirac fermions $\psi (x)$ :
\be
\cL = -{1\over {4e^2}} F_{\m\n}^2 \(\P\) + {\bar \psi} \( i{\not\!\!\pa} -
{\not\!\! A}\(\P\) -m\) \psi
\lab{mini-QED}
\ee
where $F_{\m\n} \( \P\)$ and $ A_\m \( \P\)$ are given by \rf{F-primitiv} and
\rf{A-primitiv}, respectively, {\sl i.e.}, the Lagrangian
\rf{mini-QED} describes QED with special type of composite ``photons''.
Let us stress that, although containing fourth power of derivatives on
$\P^a$, the Lagrangian \rf{mini-QED} is only second order w.r.t.
time-derivatives.

The model \rf{mini-QED}, called
``mini-QED'', appeared previously in ref.\ct{mini-QED} from a somewhat
different motivation. Unlike ordinary QED, however, we now observe that the
mini-QED model \rf{mini-QED} does not possess local gauge invariance but
rather it is invariant under infinite-dimensional {\em global}
volume-preserving-(symplecto-)diffeomorphic $\Symp{\cT^{2n}}$ Noether
sym\-metry\foot{The group parameters $\g_{\vec{n}}$ (cf. eq.\rf{torus}) are
{\em constant} space-time independent ones.} :
\br
\P^a (x) \to G^a \(\P (x)\) \approx
\P^a (x) + \om^{ab} \partder{\G \(\P (x)\)}{\P^b (x)}
\lab{Phi-transf} \\
\psi (x) \to e^{\L \(\P (x)\)} \psi (x) \qquad ,\quad
\L \(\P (x)\) \approx \G\(\P (x) \) - \h \P^a (x) \partder{\G}{\P^a}
\lab{psi-transf}
\er
Of course, the Lagrangian \rf{mini-QED} is also invariant under the usual {\em
global}
$U(1)$ symmetry: $ \psi \to e^\a \psi \; ,\; \P^a \to \P^a$ .

The above construction can be straightforwardly generalized for the higher
``$p$-form'' antisymmetric tensor gauge theories \ct{p-form}. Namely,
consider again a set of $s(\equiv p+1)$ zero-dimensional scalar fields
$\lcurl \P^a (x) \rcurl_{a=1}^s$ taking values in a smooth manifold
$\cT^s$. The pull-back of its canonical volume $s(\equiv p+1)$-form to
Minkowski space-time gives rise to an antisymmetric $s$-tensor field strength
and its associated antisymmetric $s-1$-tensor potential:
\br
\frac{1}{s!} \vareps_{a_1 \ldots a_s} d\P^{a_1} \ldots d\P^{a_s} =
\frac{1}{s!} F_{\m_1\ldots \m_s}\(\P\) dx^{\m_1} \wedge\cdots\wedge dx^{\m_s}
\lab{s-diff-form} \\
F_{\m_1\ldots \m_s} \(\P\) = \vareps_{a_1 \ldots a_s} \pa_{\m_1} \P^{a_1}\ldots
\pa_{\m_s} \P^{a_s}
\lab{F-primitiv-s} \\
F_{\m_1\ldots \m_s} \(\P\) = s! \pa_{\lb \m_1} A_{\m_2\ldots \m_s\rb} \(\P\)
\quad ,\quad  A_ {\m_1\ldots \m_{s-1}} \(\P\) =
{1\over s} \vareps_{a_1 \ldots a_s} \P^{a_1} \pa_{\m_1} \P^{a_2}\ldots
\pa_{\m_{s-1}} \P^{a_s}
\lab{A-primitiv-s}
\er
As in eq.\rf{transform}, one readily verifies that the field strength
\rf{F-primitiv-s} is invariant under arbitrary
field transformations (reparametrizations) $\P^a (x) \to G^a \(\P (x)\)$
belonging to the infinite-dimensional group $\Vol{\cT^s}$ \rf{vol-diff-group},
whereas its potential \rf{A-primitiv-s} undergoes a $\P$-dependent
$s-2$-rank local gauge transformation (cf. \rf{vol-diff-alg}--\rf{rot}) :
\br
A_ {\m_1\ldots \m_{s-1}} \(\P\) \to A_ {\m_1\ldots \m_{s-1}} \(\P\) +
(s-1)! \pa_{\lb \m_1} \L_{\m_2\ldots \m_{s-1}\rb} \(\P\)
\lab{transform-s} \\
\L_{\m_1\ldots \m_{s-2}} \(\P\) =
\llb \(  1 - {1\over s} \P^b \partder{}{\P^b}\) \G_{a_1\ldots a_{s-2}} +
\frac{(s-3)!}{s} \P^b \partder{}{\P^{\lb a_1}}
\G_{\v b\v a_2\ldots a_{s-2}\rb}\rrb \times  \nonu \\
\times \pa_{\m_1}\P^{a_1}\ldots \pa_{\m_{s-2}}\P^{a_{s-2}}
\lab{gauge-s}
\er

Now, similarly to mini-QED model \rf{mini-QED} we can construct a
``mini-Kalb-Ramond'' model (cf. \ct{K-R}) using
\rf{F-primitiv-s}--\rf{A-primitiv-s} for $s=3$ :
\be
\cS = - \frac{1}{3!} \int d^D x \, F^2_{\m\n\l} \(\P (x)\)  +
\int d^2 \s \, {1\over 3} \vareps_{abc} \P^a (x(\s ))
\pa_{\s_1} \P^b (x(\s )) \pa_{\s_2} \P^c (x(\s ))
\lab{mini-KR}
\ee
where the second integral is over the string world-sheet given by
$x^\m = x^\m (\s )$ . Again as in \rf{mini-QED}, the Lagrangian
\rf{mini-KR} being of sixth order w.r.t. $\P^a$-derivatives is only
quadratic w.r.t. time-derivatives.

\lskip
{\large {\bf 4. Equations of Motion and Symmetries}}
\mskp
First, let us consider even-dimensional space-time theories.
As already pointed out in \ct{mini-QED}, in the case when the
dimension of $\P$-target space is equal to, or greater than,
the space-time dimension, {\sl i.e.}, $dim \(\cT^{2n}\) \equiv 2n \geq D$,
any vector potential can be represented (in a non-unique way) in the form
\rf{A-primitiv}. This is not any more true when $dim \(\cT^{2n}\) \leq D-2$ ,
since then the topological density $\vareps^{\m_1 \m_2 \ldots \m_{D}}
F_{\m_1\m_2}\(\P\) \cdots F_{\m_{D-1}\m_{D}}\(\P\)$, and even the associated
topological Chern-Simons current, are identically zero :
\br
\vareps^{\m_1 \m_2 \ldots \m_{D}} A_{\m_2}\(\P\) F_{\m_3\m_4}\(\P\) \cdots
F_{\m_{D-1}\m_{D}}\(\P\) =   \nonu   \\
\h \om_{a_1 \lb a_2} \ldots \om_{a_{D-1} a_{D}\rb}
\vareps^{\m_1 \m_2 \ldots \m_{D}}
\P^{a_1} \pa_{\m_2}\P^{a_2} \pa_{\m_3}\P^{a_3} \cdots
\pa_{\m_{D}}\P^{a_{D}} = 0  \quad (~{\rm for} ~2n \leq D-2~)
\lab{CS-current}
\er
Here again the square brackets indicate total antisymmetrization of indices.
Thus, as emphasized in \ct{mini-QED}, mini-QED with a target space of the
scalar
``primitives'' satisfying the condition $dim \(\cT^{2n}\) \leq D-2$
does {\em not} exhibit the usual axial anomalies.

Similar phenomena occur in odd space-time dimensions $D$. Namely, when
$dim \(\cT^{2n}\) \geq D+1$ ~
any vector potential can be represented (in a non-unique way) in the form
\rf{A-primitiv}, whereas when $dim \(\cT^{2n}\) \leq D-1$ ~the topological
Chern-Simons term is identically zero.

The equations of motion for \rf{mini-QED} are:
\br
\pa_\m \P^a \llb {1\over {e^2}} \pa_\n F^{\m\n}\(\P\)  +
{\bar \psi}\g^\m \psi \rrb = 0
\lab{phi-motion}  \\
\( i{\not\!\!\pa} - {\not\!\! A}\(\P\) - m \) \psi = 0
\lab{psi-motion}
\er
The infinite set of Noether currents corresponding to the $\Symp{\cT^{2n}}$
symmetry of \rf{mini-QED} reads:
\br
J^\m \lb f\rb = f\(\P\) \llb {1\over {e^2}} \pa_\n F^{\m\n}\(\P\)  +
{\bar \psi}\g^\m \psi \rrb
\lab{current-1}    \\
\pa_\m J^\m \lb f\rb = 0    \quad, \qquad
\d_{\G} J^\m \lb f\rb = J^\m \lb \pbbr{f}{\G}\rb
\lab{current-2}
\er
where $f\(\P\)$ is an arbitrary (smooth) function on the $\P$-target space
$\cT^{2n}$ and $\d_\G$ indicates an infinitesimal symplecto-diffeomorphic
transformation \rf{Phi-transf}--\rf{psi-transf}.
In the particular case of torus target space \rf{torus} we have:
\be
J^\m \lb f\rb = \sum_{\vec{n}} f_{\vec{n}} J_{\vec{n}}^\m \quad ,\quad
J_{\vec{n}}^\m = e^{i\vec{n} .\vec{\P}}\llb {1\over {e^2}}\pa_\n F^{\m\n}\(\P\)
+ {\bar \psi}\g^\m \psi \rrb
\lab{current-3}
\ee

The first thing one notices in eqs.\rf{phi-motion}--\rf{psi-motion} is that
the $\P$-equations of motion differ from the usual Maxwell equations (with a
``composite'' vector potential) just by the contraction of the latter with the
$2n \times D$ matrix $\Vert \pa_\m \P^a \Vert$ . This observation
together with eq.\rf{CS-current} lead us to the following:
\mskp
{\bf Main statement (even ${\bf D}$).} {\em The $\P$-equations of motion
\rf{phi-motion} coincide with the Maxwell ones if and only if
$rank \Vert \pa_\m \P^a \Vert \geq D$ (cf. eq.\rf{CS-current}).
This implies that both the topological Chern-Simons
current as well as its divergence (the topological density) are non-zero.
Alternatively, the $\P$-equations of motion
\rf{phi-motion} possess non-Maxwell solutions if either the topological
Chern-Simons current or the topological density are zero, i.e.,
$rank \Vert \pa_\m \P^a \Vert \leq D-2$ (cf. eq.\rf{CS-current}).}
\mskp
Completely analogous statement holds for odd space-time dimensions $D$ upon
substituting $D \to D+1$ above.
\mskp
Therefore, in the sequel we shall be interested in the second case,
{\sl i.e.}, the case with zero topological Chern-Simons current, which
implies the existence of a non-Maxwell sector of mini-QED \rf{mini-QED}.
In particular, when $dim \( \cT^{2n}\) \equiv 2n \leq D-2$ ~mini-QED is
generically non-Maxwell. It can be viewed as a truncation of a model
proposed in \ct{Guend-Owen} describing ordinary Maxwell $D=4$ QED coupled to a
pseudo-scalar ``pion'' field $\p$ :
\be
\cL (\psi, A;\p ) = -{1\over {4e^2}} F_{\m\n}^2 +
{\bar \psi} \( i{\not\!\!\pa} - {\not\!\! A} -m\) \psi +
\h \p \vareps^{\m\n\k\l} F_{\m\n} F_{\k\l} + \cL (\p )
\lab{Guend-Owen}
\ee
where $\cL (\p )$ contains the kinetic and self-interacting terms for the
``pion''. Ignoring in \rf{Guend-Owen} the purely ``pionic'' Lagrangian,
{\sl e.g.}, considering the heavy mass limit for $\p$, we see that the field
$\p$ becomes a non-dynamical Lagrange multiplier enforcing the constraint for
zero topological density. Thus, mini-QED is a particular realization of the
heavy mass limit of the model \rf{Guend-Owen} with the additional
condition for a vanishing topological Chern-Simons current.
\lskip
\noindent
{\large {\bf 5. Physical Consequences: New Effects }}
\mskp
For definiteness,
from now on we shall consider mini-QED \rf{mini-QED} in ordinary $D=4$
space-time and with a two-dimensional $\P$-target space.% , {\sl e.g.} a torus.
\mskp

{\bf 5.1 Magneto-Hydrodynamical Analogy}
\sskp
The existence of non-Maxwell solutions to \rf{phi-motion} means that there
exists a non-zero additional current ${\cJ}^\m \(\P\)$ --
suitable functional of $\P^a$, such that:
\be
{1\over {e^2}} \pa_\n F^{\m\n}\(\P\)  + {\bar \psi}\g^\m \psi +
{\cJ}^\m \(\P\) = 0 \quad , \quad  \pa_\m \P^a {\cJ}^\m \(\P\) = 0
\lab{phi-J}
\ee
The second eq.\rf{phi-J} implies also:
\be
A_\m \(\P\) {\cJ}^\m \(\P\) = 0 \quad ,\quad F_{\l\m}\(\P\) {\cJ}^\m \(\P\) = 0
\lab{A-F-J}
\ee
Using the standard three-dimensional Maxwell notations, one can rewrite the
second eq.\rf{A-F-J} as:
\br
\vec{E}\(\P\).\vec{\cJ}\(\P\) = 0
\lab{MHD-0} \\
- {\cJ}^0 \(\P\) \vec{E}\(\P\) + \vec{B}\(\P\) \times \vec{\cJ}\(\P\) = 0
\lab{MHD-vec}
\er
{}From \rf{MHD-vec} one easily verifies that non-zero solutions for
${\cJ}^\m \(\P\)$ exist if  $\vec{E}\(\P\).\vec{B}\(\P\) =0$
which is precisely the expression for zero topological density
$\vareps^{\m\n\k\l} F_{\m\n}\(\P\) F_{\k\l}\(\P\)$ . Then, eq.\rf{MHD-vec}
can be represented in the following ``magneto-hydrodynamical'' form
(cf. \ct{L-L}) :
\be
\vec{E}\(\P\) + \frac{\vec{\cJ}\(\P\)}{{\cJ}^0 \(\P\)} \times \vec{B}\(\P\) = 0
\lab{MHD}
\ee
where $\vec{v}\(\P\) \equiv \vec{\cJ}\(\P\)/{\cJ}^0 \(\P\)$ can be viewed as
a $\P$-dependent velocity field. For such velocity, however,
$\mid \vec{v} \mid$ may exceed $1$. In any case, the condition
$\vec{E}\(\P\).\vec{B}\(\P\) =0$ guarantees that there is always a frame
where either the electric field is zero (as in usual
magnetohydrodynamics) or where the magnetic field is zero (``dual
magnetohydrodynamics'').
\mskp

{\bf 5.2 Electromagnetic Vacuum in Mini-QED}
\sskp
Now, let us consider the mini-QED electron dynamics in the presence of
electromagnetic vacuum:
\be
F_{\m\n}\(\P_{vac}\) = 0
\lab{el-magn-vac}
\ee
The general solution to \rf{el-magn-vac} can be taken in the form:
$\P_{vac}^1 (x) \equiv u(x) \; ,\; \P_{vac}^2 (x) = B(u(x))$ -- arbitrary
local function of $u$.
Then eqs.\rf{psi-motion}--\rf{phi-motion} become (after an
appropriate phase transformation of the Dirac field) :
\br
\( i{\not\!\!\pa} - m \) \psi_{vac} = 0  \\
\lab{psi-vac-motion}
\({\bar \psi}_{vac}\g^\m \psi_{vac} \) \pa_\m u  = 0
\lab{phi-vac-motion}
\er
The last equation means that $\P_{vac}^a (x)$ are constant along the
straight free-electron world-lines, {\sl i.e.},
$\partder{}{\t} \P_{vac}^a (x(\t )) = 0$ .
Thus, one gets usual plain-wave solutions
without the self-energy problem of ordinary QED (where one would get that the
current must vanish simultaneously).
\mskp

{\bf 5.3 Gauge-Invariant Photon Mass Generation}
\sskp
Let us now show that mini-QED built out of two primitive scalars
allows propagation of massive gauge-invariant modes in $2+1$ and in $3+1$
dimensions.

First, let us consider the $2+1$-dimensional case.
The basic observation is that the equations of motion for mini-QED in
the absence of sources:
\be
\pa_\m \P^a \,  \pa_\n F^{\m\n}\(\P\) = 0
\lab{free-mini-QED}
\ee
allow, apart from the pure
Maxwell solutions where $ \pa_\n F^{\m\n}\(\P\)  = 0 $, also non-Maxwell
solutions of the following form:
\be
\pa_\n F^{\m\n}\(\P\)  =  \mu \(\P\) F^{\ast \m} \(\P\) \quad ,\qquad
F^{\ast\m} \(\P\) = \h\vareps^{\m\n\a}F_{\n\a} \(\P\)
\lab{massive-1}
\ee
Consistency of \rf{massive-1} with the ``free'' mini-QED eqs.\rf{free-mini-QED}
is easily verified remembering the form of $F^{\m\n} \(\P\) = \vareps_{ab}
\pa_\m \P^a \pa_\n \P^b$ . Indeed, substituting \rf{massive-1} in the l.h.s.
of \rf{free-mini-QED} one gets:
\be
\pa_\m \P^a \, \pa_\n F^{\m\n}\(\P\) =
\h \m (\P ) \vareps^{\m\n\a}F_{\n\a} \(\P\) \pa_\m \P^a =
\h \m \(\P\)\vareps^{\m\n\a} \vareps_{bc} \pa_\m \P^{\lb a}
\pa_\n \P^b \pa_\a \P^{c\rb} = 0
\lab{consist-1}
\ee
where the last equality follows from antisymmetrization of three indices
taking only two values.

Furthermore, taking the divergence of eq.\rf{massive-1} on
 both sides we obtain:
\be
\pa_\m \pa_\n F^{\m\n}\(\P\) = 0
= \partder{\m \(\P\)}{\P^a} \,\pa_\m \P^a F^{\ast\m} \(\P\)
\lab{consist-2}
\ee
This is consistent since, as we showed in \rf{consist-1},
$ \pa_\m \P^a F^{\ast\m} \(\P\) = 0 $. In the case where
$ \mu \(\P\) = \m =const $ ,
eq.\rf{massive-1} coincides with the equation of the topologically massive
Chern-Simons theory (see \cite{PV-1,PV-2} and references therein). However,
the solution \rf{massive-1} to mini-QED eqs.\rf{free-mini-QED} exhibits an
interesting new feature, namely that here the mass is not determined,
{\sl i.e.}, it could have any value.

The solutions of eq.\rf{massive-1} for the case $\mu \(\P\) = \m =const $
are known \cite{PV-2}. As it has been shown in ref.\ct{PV-2},
the field strengths $ F_{0i}= E_{i}$ and
$ B = \vareps_{ij} \pa_i A_{j} $ can be expressed in terms of a massive free
field $\varphi$ satisfying $( \Box + \m^2) \varphi =0 $ according to:
 \be
 E_{i} = \vareps_{ij}\hat{\pa}_{j}\dot{\varphi}
 + \mu \hat{\pa}_{j}\varphi \quad ,\quad
 B = - \sqrt{-\nabla^2} \varphi \qquad, \quad
 \hat{\pa}_{j} = \frac{{\pa}_{j}}{\sqrt{-\nabla^2}}
 \lab{massive-2}
 \ee
where the dot indicates $\pa_0$-derivative and $i,j =1,2$ .

It is of interest to see how the solutions \rf{massive-2} can be
 expressed in terms of the primitive scalar fields $\P^a$, {\sl i.e.},
$E_i \(\P\) = \vareps_{ab} {\dot \P}^a \pa_i \P^b $ and
$B\(\P\)= \h \vareps^{ij} \vareps_{ab} \pa_i \P^a \pa_j \P^b $ .
For this purpose, let us consider $ E_{i}$ and $ B $ \rf{massive-2} given in
terms of $\varphi$ which is a simple plane wave of the form
$ \varphi = A \cos \(k^{0}x^{0}- k^{i}x^{i}\) $, namely:
\br
\hat{\pa}_{i}\varphi= - A\hat{k^{i}}\sin \(k^{0}x^{0}- k^{i}x^{i}\)
\quad ,\quad
\hat{\pa}_{i} \dot{\varphi}
= - A\hat{k^{i}}k^{0}\cos \(k^{0}x^{0}- k^{i}x^{i}\)  \nonu \\
B = - A\sqrt{{\vec{k}}^2} \cos \(k^{0}x^{0}- k^{i}x^{i}\)
\lab{plane}
\er
Let us consider the rest frame, that is the limit $ k^{i} \longrightarrow 0 $.
In such a case we obtain (since $k^{0} \longrightarrow  \m $) :
 \be
 E_{i} = -A \m \llb \vareps_{ij}{\hat k}_{j} \sin (\mu t) +
 {\hat k}_{i} \cos (\mu t) \rrb \quad , \quad B = 0
 \lab{massive-3}
 \ee

 We see that when $ k^{i} \longrightarrow 0 $ the unit vector $ \hat{ k_{i}} $
 does not disappear from the expression for the electric field in the rest
 frame \rf{massive-3}. Such a unit vector $ \hat{ k_{i}} $ can be arbitrarily
chosen in any space direction. For example,
 choosing $ \hat{ k_{i}} $ in the $x$-direction implies that
 $\epsilon_{ij}\hat{ k_{j}}$ is in the $y$-direction. Therefore,
the electric field described by eq.\rf{massive-3} is a rotating one in the
$x-y$ plane. The effect of choosing $ \hat{ k_{i}} $ along a different space
direction is equivalent to a time shift in eq.\rf{massive-3}.

 A choice for $\P^{1} $ and $\P^{2} $ that reproduces the configuration
 \rf{massive-3} is:
 \be
 \P^{1} = -t\m \quad ,\quad
 \P^{2} = A x^{i} \llb \vareps_{ij}\hat{k_{j}} \sin (\mu t) +
 \hat{ k_{i}} \cos(\mu t) \rrb
  \lab{massive-4}
 \ee
The generalization of \rf{massive-4} to an arbitrary coordinate system where
the $\varphi$-``particle'' has an arbitrary momentum reads:
 \be
 \P^{1} = - k_{\n}x^{\n} \quad , \quad
\P^{2} = x_{\n} \llb\vareps^{\n\a\b}s_{\a} k_{\b} \sin (k_{\s}x^{\s} )
+ \m s^{\n} \cos (k_{\s}x^{\s}) \rrb
 \lab{massive-5}
\ee
where $ s^{\n} $ is any vector satisfying $ s^{\n} k_{\n} = 0 $, {\sl i.e.},
it is such that in the rest frame $ s^{\n} = \( 0, \vec{s}\) $. Thus,
eqs.\rf{massive-5} provide a solution to the sourceless mini-QED equations
of motion \rf{free-mini-QED} describing gauge invariant massive photons with
an arbitrary mass.

The $2+1$-dimensional non-Maxwell mini-QED solution given above can be
easily embedded in the $3+1$-dimensional mini-QED given by \rf{free-mini-QED}.
In this case we have to select arbitrarily one space dimension which we do
not want to appear in the solution, let us say $z$ . Then, instead of
\rf{massive-1} we should use:
\be
\pa_\n F^{\m\n}\(\P\)  =  \h \mu \(\P\) \vareps^{3\m\n\a} F_{\n\a}\(\P\)
\lab{massive-6}
\ee
where there is no dependence of any field on $z$ .

The physical consequences of a continuous mass spectrum, similar to the one
found here, have been discussed in ref.\ct{Coy-Wu}.

\lskip
\noindent
{\large {\bf 6. Open Problems and Outlook; Mini-QED as an Induced Massive
Gravity}}
\mskp

The next most important question is quantization of the mini-QED model
\rf{mini-QED}. Applying formally the general arguments of renormalization
theory \ct{Renorm}, namely that the possible counterterms canceling the
ultra-violet divergences must be local, covariant functionals of the
constituent fields of dimension $D=4$ and {\em preserving the global Noether}
$\Symp{\cT^{2n}}$ {\em symmetry}, one easily deduces that these counterterms
will be exactly the same as in ordinary QED, {\sl i.e.}, the renormalized
mini-QED Lagrangian will be of the same form as \rf{mini-QED} with the usual
charge-, mass- and multiplicative field renormalizations.

The problem with
\rf{mini-QED} is, however, that the kinetic term for the scalar ``primitives''
$\P^a$ is of a non-standard quartic form which does not allow to develop
the usual Feynman diagram expansion. Moreover, if we would add
to \rf{mini-QED} an
ordinary kinetic term for $\P^a$, the latter would break the $\Symp{\cT^{2n}}$
symmetry. Obviously, we have to rewrite \rf{mini-QED} in an equivalent form
with
the help of auxiliary fields which would bilinearize the $\P^a$-part of the
Lagrangian --- much in the same way one deals with models having four-fermion
interaction terms.

Considering again for simplicity two-dimensional $\P$-target space and using
the
simple identity:
\be
\vareps_{ab} \vareps_{cd} = \d_{ac}\d_{bd} - \d_{ad}\d_{bc}
\lab{ident}
\ee
one can represent the first ``Maxwell'' term in \rf{mini-QED} as follows:
\be
F_{\m\n}^2 \(\P\) = - \( \pa_\m \P^a \pa_\n \P_a\)
\llb \eta^{\m\k} \eta^{\n\l} - \eta^{\m\n} \eta^{\k\l} \rrb
\( \pa_\k \P^b \pa_\l \P_b\)
\lab{F-2}
\ee
where $\eta^{\m\n}$ is the flat Minkowski space-time metric tensor.
Taking into account \rf{F-2}, the mini-QED Lagrangian \rf{mini-QED} acquires
the
following equivalent form:
\be
\cL = \h h^{\m\n} \pa_\m \P^a \pa_\n \P_a -
\frac{e^2}{4} h^{\m\n}\(\eta^{\m\k}\eta^{\n\l} -
{1\over 3}\eta^{\m\n}\eta^{\k\l}\)  h^{\k\l} -
\({\bar \psi}\g^\m \psi\) \h \vareps_{ab} \P^a \pa_\m \P^b  +
{\bar \psi} i{\not\!\!\pa} \psi
\lab{mini-QED-h}
\ee
Here $h^{\m\n}$ is an auxiliary symmetric tensor field classically given by:
\be
h^{\m\n} = \frac{1}{e^2} \llb \pa_\m \P^a \pa_\n \P_a -
\eta^{\m\n} \(\pa_\l \P^b \pa^\l \P_b\) \rrb
\lab{h-mn}
\ee
The first term in eq.\rf{mini-QED-h} indicates that $h^{\m\n} \equiv
\sqrt{-g} g^{\m\n}$ can be viewed as composite gravitational field (cf.
\rf{h-mn}) coupled to the scalar primitives $\P^a$ and being massive (cf. the
second term in \rf{mini-QED-h}) with a mass proportional to the square of
the electric charge. Integrating out the $\P^a$-fields will produce
(upon neglecting the fermions) the standard quantum effective action of
scalar fields in a curved background \ct{Bir-Dav}
with the usual divergent cosmological, Einstein- and $R^2$-terms for
$h^{\m\n} \equiv \sqrt{-g} g^{\m\n}$ . Thus, we find an intriguing
equivalence of the mini-QED model \rf{mini-QED} with a theory of purely
induced {\em and} massive gravity \rf{mini-QED-h} which, however, is formally
non-renormalizable. On the other hand, unlike ordinary
gravity (with dimensionless $h^{\m\n} \equiv \sqrt{-g} g^{\m\n}$ )
the field $h^{\m\n}$ in \rf{mini-QED-h} has dimension 2 (cf. \rf{h-mn}) and,
therefore,
the usual divergent non-renormalizable semi-classical expansion of ordinary
gravity interacting with matter fields around flat background metric
is not appropriate for mini-QED in
the form \rf{mini-QED-h}. Obviously, a bilinearization of mini-QED
\rf{mini-QED} different from \rf{mini-QED-h} is needed for its proper
quantization -- a question, which is currently under investigation.

Another possible venue is to exploit the Ward identities for the
infinite-dimensional global Noether $\Symp{\cT^{2n}}$ symmetry
\ct{shorty} to obtain
non-perturbative information for the quantum correlation functions of
mini-QED. In particular, an interesting important question arises about
possible quantum deformation of the classical Noether $\Symp{\cT^{2n}}$
symmetry, {\sl e.g.}, the area-preserving-diffeomorphisms'
$\Symp{\cT^{2}}$ symmetry might acquire central extension \rf{ext-torus-alg}
and/or be deformed into (centrally extended) $\Win1$-symmetry, as it happens
in $D=2$ conformal field theory (cf., {\sl e.g.} \ct{W-inf}).
\lskip
\small
{\bf Acknowledgements.}
E.N. and S.P. gratefully acknowledge support from the Ben-Gurion University,
Beer-Sheva. E.N. is also thankful to H. Aratyn for cordial hospitality at
the University of Illinois at Chicago where this work was completed.

\end{document}